\documentclass{llncs}
\usepackage{makeidx}  
\usepackage{graphicx}
\usepackage[]{graphicx}
\usepackage{amsmath }
\usepackage{color}
\begin{document}
\mainmatter              

\title{Cross-Domain \\ Conditional Generative Adversarial Networks for\\ Stereoscopic Hyperrealism in Surgical Training} 

\titlerunning{Stereoscopic Hyperrealim in Surgical Training}  
%
\author{Sandy~Engelhardt\inst{1,3}\and
Lalith Sharan\inst{1,3} \and
Matthias~Karck\inst{2}\and
Raffaele~De~Simone\inst{2} \and
Ivo~Wolf\inst{1}}
\authorrunning{Engelhardt et al.} 
%
%
\institute{
Faculty of Computer Science, Mannheim University of Applied Sciences, Germany \\
\email{s.engelhardt@hs-mannheim.de}
\and
Department of Cardiac Surgery, Heidelberg University Hospital, Germany
\and
Dep. of Simulation and Graphics, Magdeburg University, Germany}

\maketitle              

\begin{abstract} 
Phantoms for surgical training are able to mimic cutting and suturing properties and patient-individual shape of organs, but lack a realistic visual appearance that captures the heterogeneity of surgical scenes. In order to overcome this in endoscopic approaches, hyperrealistic concepts have been proposed to be used in an augmented reality-setting, which are based on deep image-to-image transformation methods. 
Such concepts are able to generate realistic representations of phantoms learned from real intraoperative endoscopic sequences. Conditioned on frames from the surgical training process, the learned models are able to generate impressive results by transforming unrealistic parts of the image (e.g.\ the uniform phantom texture is replaced by the more heterogeneous texture of the tissue). Image-to-image synthesis usually learns a mapping $G:X~\to~Y$ such that the distribution of images from $G(X)$ is indistinguishable from the distribution $Y$. However, it does not necessarily force the generated images to be consistent and without artifacts. In the endoscopic image domain this can affect depth cues and stereo consistency of a stereo image pair, which ultimately impairs surgical vision. We propose a cross-domain conditional generative adversarial network approach (GAN) that aims to generate more consistent stereo pairs. The results show substantial improvements in depth perception and realism evaluated by 3 domain experts and 3 medical students on a 3D monitor over the baseline method. In 84 of 90 instances our proposed method was preferred or rated equal to the baseline.



\keywords{Generative adversarial networks, minimally-invasive surgical training, augmented reality, mitral valve simulator, laparoscopy}
\end{abstract}
\section{Introduction}
Minimally invasive surgery is characterized by a restricted view of the surgical target. In many such procedures, the only way to observe the surgical field is with endoscopic vision on an external display, which is commonly associated with an impaired depth perception. This situation requires an excellent hand-eye coordination, exceptional skills and dexterity with instruments, which should be trained with surgical simulators before performing it on patients. On such simulators, the photo-realistic fidelity and depth perception is a key feature that can improve the transfer ratio of trainees to real surgeries. However, most virtual or physical simulators lack such properties, due to limited capabilities of modelling realistic textures or by current material that is used for surgical simulation. 

In our previous work \cite{EngelhardtMICCAI}, a deep learning-based concept to tackle the issue of photo-realism of surgical simulations was presented. The approach was coined \textit{hyperrealism}, which is able to map patterns learned from intraoperative video sequences onto the video stream captured during simulated surgery on anatomical replica. Used within an augmented reality setting, \textit{hyperrealism} is defined as a new augmented reality paradigm on the Reality-Virtuality continuum \cite{milgram_taxonomy_1994}, which is closer to `full reality' in comparison to other concepts where artificial overlays are superimposed on a video frame. In a hyperrealistic surgical training environment, the parts of the simulated environment that look unnatural are replaced by realistic appearances. Parts that already look natural ideally stay the same. 
It is shown that such an approach greatly improves reproduction of the intraoperative appearance during training and therefore makes minimally-invasive surgical training more realistic. 
The approach is in principle also employable for enhancing photo-realism of virtual training simulators, as shown by Luengo et al. \cite{Luengo2018SurRealES} who used a deep learning approach that relies on style transfer.

Methodologically, our  proposed approach is based on so-called unpaired deep image-to-image transformation methods \cite{Zhu2017}. The underlying concept is to use adversarial training of a generator and a discriminator network, and to employ a cycle between the two input domains to generate realistic looking images. The key to the success of such generative adversarial networks (GANs) is the idea of an adversarial loss that forces the generated images to be, in principle, indistinguishable from real images. Such concepts are able to generate realistic representations of phantoms learned from real intraoperative endoscopic sequences \cite{EngelhardtMICCAI}. Conditioned on frames from the surgical training process, the learned models are able to generate impressive results by transforming unrealistic parts of the image (e.g.\ the uniform phantom texture is replaced by the more heterogeneous texture of the tissue). 
However, the traditional CycleGAN approach \cite{Zhu2017} neither enforces temporal coherence, which was incorporated in our previous contribution \cite{EngelhardtMICCAI}, nor enforces a stereo pair to be consistent. 

A recent work published at CVPR 2018 \cite{Chen_2018_CVPR} was the first to address stereoscopic neural style transfer. Approaches beforehand only dealt with monocular style transfer. The authors showed that independent application of stylization approaches to left and right views of stereoscopic images does not preserve the original disparity consistency in the final stylization results, which causes 3D fatigue to the viewers on a 3D display \cite{Chen_2018_CVPR}. Chen et al. \cite{Chen_2018_CVPR} incorporated a disparity loss into the style loss function that is employable in virtual scenes with known disparity information. In this contribution, we want to tackle the same issue for physical surgical simulators, i.e.\ generation of more consistent stereo-images without relying on ground truth disparity, which is more complicated to obtain in the medical domain. In order to achieve this, a novel cross-domain conditional GAN is introduced in the following. 


\begin{figure}[t]
   \begin{center}
 \includegraphics[width=\linewidth]{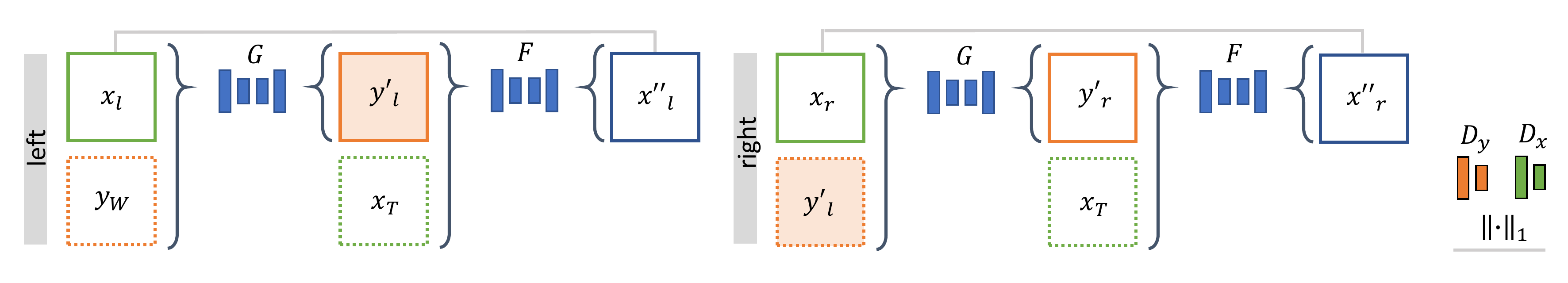}
   \end{center}
   \caption[] 
   { \label{fig:Architecture} Proposed architecture that shows the $X$$\to$$Y$$\to$$X$-cycle for a stereo pair $(x_l,x_r)$. In contrast to classical generators, each generator $G$ and $F$ takes two inputs and generates one output. The second input image, e.g. $y_W$, $x_T$, is taken from the other domain and can be chosen randomly. To enable a better consistency, the output of $G$, which is $y'_l$, is chosen as a second input in the generation cycle of the right image. 
   Discriminators $D_y$ and $D_x$ evaluate real and fake images marked in orange and green. }
\end{figure}

\section{Methods}
Given unpaired training samples in two image domains $X$ and $Y$, the CycleGAN model proposed by Zhu et al.\ \cite{Zhu2017} learns a mapping (generator) $G:X \to Y$ and the reverse mapping $F:Y \to X$.
These generators are trained to produce outputs that are indistinguishable from real images in the respective target domains for discriminator networks $D_Y$ and $D_X$.
Additionally, the consistency of the cyclic mappings $F(G(x))$ and $G(F(y))$ with the respective inputs $x\in X$ and $y \in Y$ is enforced by using the  $L_1$-norm.
The proposed method builds upon this idea to learn a style transfer between an image stream from the source domain $X$ of \textit{surgical simulation} to a target domain $Y$ of \textit{intraoperative surgeries} and vice versa in the absence of paired endoscopic image samples.


\subsection{Cross Domain Conditional GAN}
For our task, the forward generator $G:X \to Y$ of the standard CycleGAN tended to create unrealistic colors and artifacts in the generated intraoperative scenes.
To overcome this, we introduce \textit{cross domain conditional GANs}.

Traditional \textit{conditional GANs} \cite{Mirza2014} learn a mapping from an observed input domain $X$ and random noise $Z$ to the output domain $Y$, $G : X \times Z \to Y$.
Isola et al. \cite{Isola2016} found the additional random noise vector to be ineffective and dropped it from the paired image translation predecessor of the unpaired CycleGAN as well as from the CycleGAN itself (in contrast to the concurrent DualGAN \cite{Yi2017}).

We propose to re-introduce an additional input, but to use a sample $y$ from the target domain distribution $p_Y$ instead of random noise to guide the training of the generator $G : X \times Y \to Y$ to realistic coloring and preservation of detail (and analogously $F:Y \times X \to X$).

For our stereo image translation task, we use a random sample $y_W \sim p_Y$ for the translation of the left image $x_l$ of the stereo pair and the generated output of the generator $y'_l:=G(x_l,y_W)$ for generating the right image $y'_r:=G(x_r,y'_l)$ to additionally support the generation of consistently colored stereo pairs, see Figure~\ref{fig:Architecture}.  

\subsection{Network Architectures}
The used network architectures of the generators and discriminators are largely the same as in the original CycleGAN approach \cite{Zhu2017}.
A TensorFlow implementation provided on GitHub\footnote{https://github.com/LynnHo/CycleGAN-Tensorflow-PyTorch-Simple}
was used as the basis and extended. All discriminators take the complete input images, which is different from the $70 \times 70$ PatchGAN approach \cite{Zhu2017}.
For the generators, 7 instead of 9 residual blocks are used, because experiments on our data showed better results for this configuration. Moreover, the generators were changed to handle a 6-channel input. 

\section{Evaluation}

A minimally invasive mitral valve repair simulator (MICS MVR surgical simulator, Fehling Instruments GmbH \& Co. KG, Karlstein, Germany) was extended with patient-specific silicone mitral valves. In comparison to other simulators, the valve replica consist of all anatomical parts, i.e. the annulus, the leaflets, the chordae tendineae and the papillary muscles. Details on the valve model production are elaborated on in a previous work \cite{Engelhardt_IJCARS}. 

An expert segmented pathological mitral valves on the end-systolic time step from echocardiographic data, which are represented as virtual models. From these models, 3D printable molds and suitable prosthetic rings were automatically generated and 3D-printed. Subsequently, 15 silicone valves were manufactured that could be anchored in the simulator onto a custom valve holder. We asked different experts and trainees to perform mitral valve repair techniques (annuloplasty, triangular leaflet resection, neo-chordae implantation) on these valves and recorded the endoscopic video stream \cite{Engelhardt_ICVTS}.

\begin{figure}[t]
   \begin{center}
 \includegraphics[width=\linewidth]{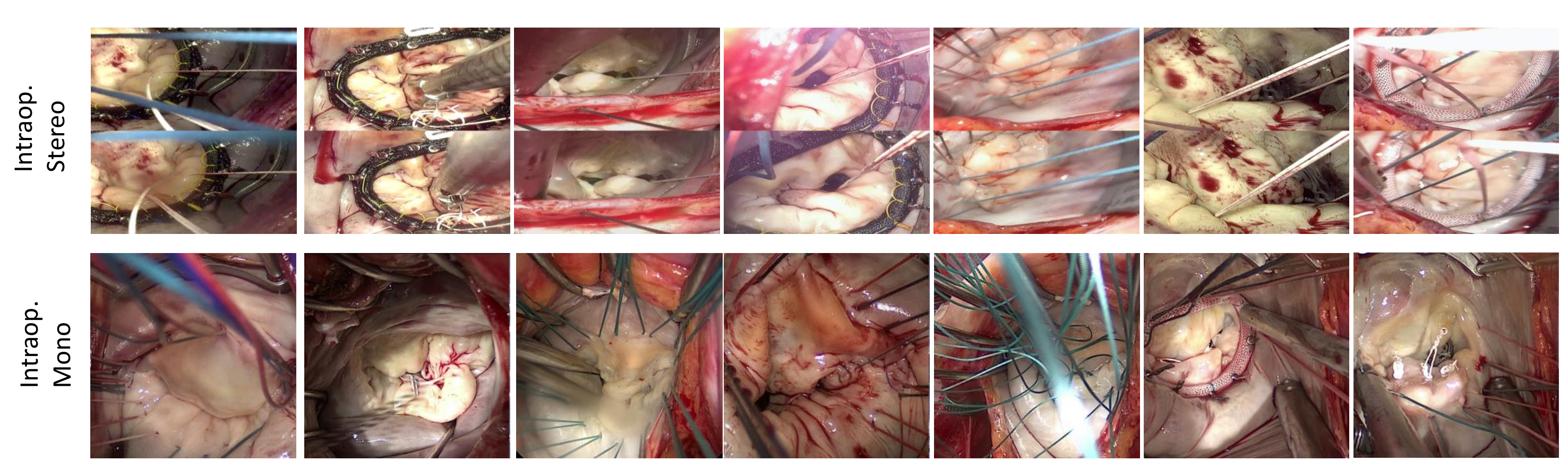}
   \end{center}
   \caption[] 
   { \label{fig:Real} Mono- and stereoscopic examples from mitral valve repair. The scenes are diverse: with or without prosthetic ring, sutures, instruments and needles, blood etc.}
\end{figure} 

\subsection{Data and Training of Network}

In total, approx.\ 240,000 stereo pair frames from the surgical training procedures were captured in full HD resolution or larger, which sums up to 9h of video material. Most of the videos were captured at 25 fps. Due to change in recording equipment, a subset of approx.\ 20,000 stereo frames was acquired at 1 fps. The training data for the network was sampled every 240th frame or every 40th frame, respectively. In total, the network training set consists of 1400 stereo pair frames from the training with the surgical phantom. To avoid overfitting of the model, valve replica shown in videos for network training were not used for network testing. 

Intraoperatively, more than 620,000 stereo pairs were captured during three minimally invasive mitral valve repair surgeries. The  frame rate varied between 60 fps and 25 fps. Scenes where the valve itself was not visible were neglected. 
For network training, a stereo pair after each 120th or 240th frame was sampled retrospectively from these videos, which sums up to approx.\ 1200 stero pairs for network training. 
The scenes are highly diverse, as the valve's appearance drastically changes over time (e.g. due to cutting of tissue, implanting sutures and prostheses, fluids such as blood and saline solution), see Fig. \ref{fig:Real}. Furthermore, occlusions or lens fogging often disturbed the recording. 

Besides the just described stereo data collection, we pre-trained our network on a monoscopic data base. The strength of our proposed method is that, it does not solely rely on a stereo pair as input, but can be also trained un-stereo-paired. The monoscopic data was put together from recordings during four open mitral valve surgeries with a monocular endoscope, in which case fewer lens occlusions and less fogging occurred. 
For the phantom recordings, half of the frames used for the monoscopic pre-training are also represented in the stereo data set.
In total, the source and target domain consisted of approx.\ 1500 single frames each.   

All monoscopic frames or each left and right image of the pair were randomly cropped and re-scaled to 256 $\times$ 512. Further data augmentation was performed by random horizontal flipping and intensity re-scaling. For all the experiments, the consistency loss was weighted with $\lambda = 20$. The Adam solver with a batch size of 1 and a learning rate of 0.0001 without linear decay to zero was used. Similar to Zhu et al. \cite{Zhu2017}, the objective was divided by 2 while optimizing $D$, which slows down the rate at which $D$ learns relative to $G$. Discriminators are updated using a history of 50 generated images rather than the ones produced by the latest generative networks \cite{Zhu2017}. 
The CycleGAN network was pre-trained for 40 epochs on the monoscopic data and then trained on the left image of the stereo pair for another 40 epochs. Similarly, our proposed network was trained on 40+40 epochs. For using the proposed network in the monoscopic case, $y_W$ is randomly chosen from the other domain in the $X$$\to$$Y$$\to$$X$ cycle and $x_W$ accordingly in the reverse cycle.

\subsection{Evaluation}

The most important factors for the proposed application are related to perception. Therefore, we first evaluated whether three domain experts (cardiac surgeons who each at least assisted mitral valve repair surgeries) and three non-experts (medical students) are able to perceive depth on a 3D monitor on interlaced stereo pairs. Secondly, we asked the surgeons how real the generated intraoperative stereo frames appear from their experiences. All answers had to be given on a 5-point Likert Scale, with 5 being the answer with the highest agreement. Related to this, we found it crucial to ask clinical questions to the domain experts in order to show the reliability of the transformation, which is associated with a change in appearance of the scene. Reliability of the transformation requires the scene to not change too drastically, meaning that neither the shape of objects  should be altered, nor additional parts should be added or taken away. The surgeons were asked 1) to diagnose the pathology of the presented valve, 2) to name the surgical instrument visible in the scene, and 3) to state which phase of the surgery is presented.
We evaluated these questions by extracting 15 random samples from our test set; each sample was shown in interlaced format on a 3D monitor. For each frame, the corresponding result from the original CycleGAN was shown directly afterwards, therefore enabling a direct comparison between our results and the baseline. 
At the start of the experiments we asked the participants of the study to rate the depth for two realistic stereo frames from the surgical phantom, in the following referred to as \textit{Test1} and \textit{Test2} example.

\section{Results}

\begin{figure}[t]
   \begin{center}
 \includegraphics[width=\linewidth]{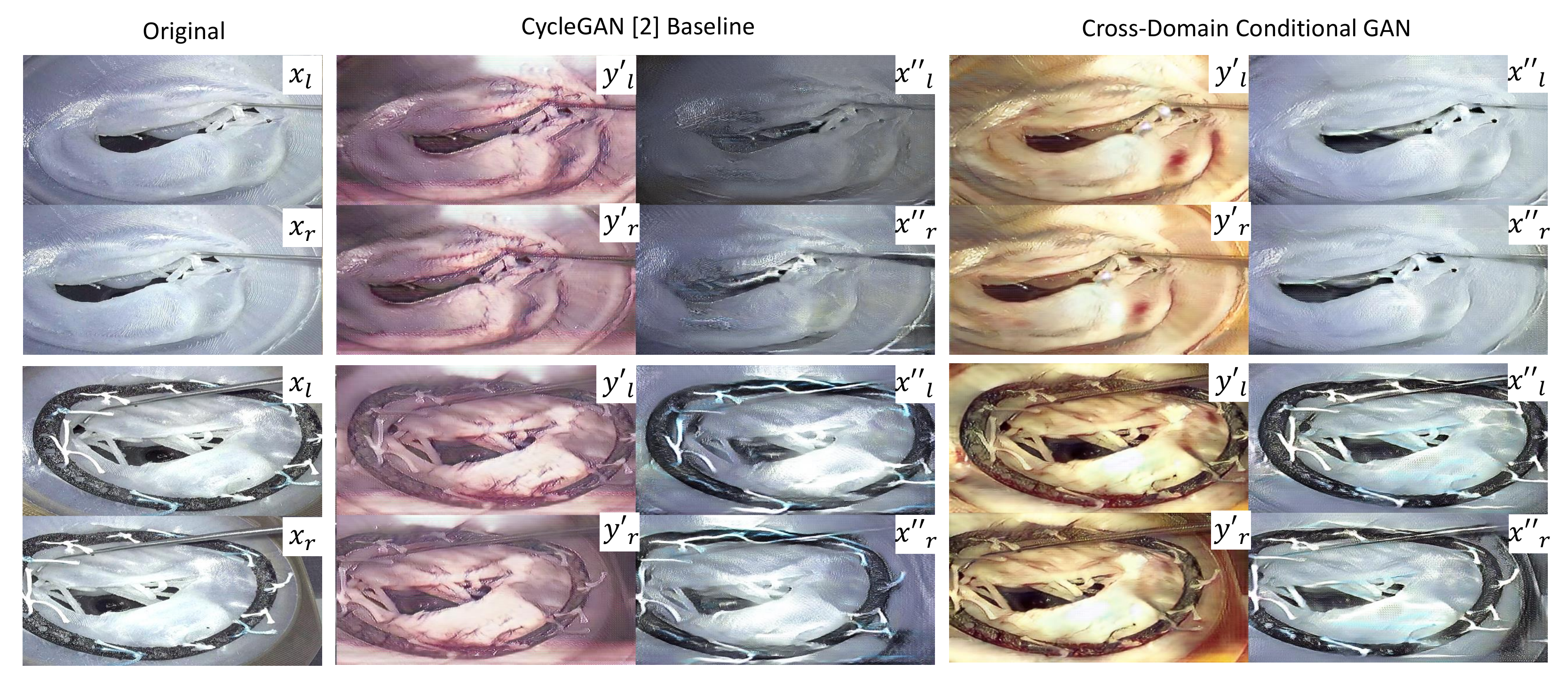}
   \end{center}
   \caption[] 
   { \label{fig:VSBaseline} Examples from CycleGAN baseline \cite{Zhu2017} and our proposed method. }
\end{figure}

Example results of our method in comparison to the baseline CycleGAN are provided in Fig. \ref{fig:VSBaseline}. A rough visual analysis shows that the structure of the valves and of the instruments are better preserved in our method.  Furthermore, the left and right image of the stereo pair appeared to be very consistent. The same was confirmed by the user study.

\begin{figure}[t]
   \begin{center}
 \includegraphics[width=\linewidth]{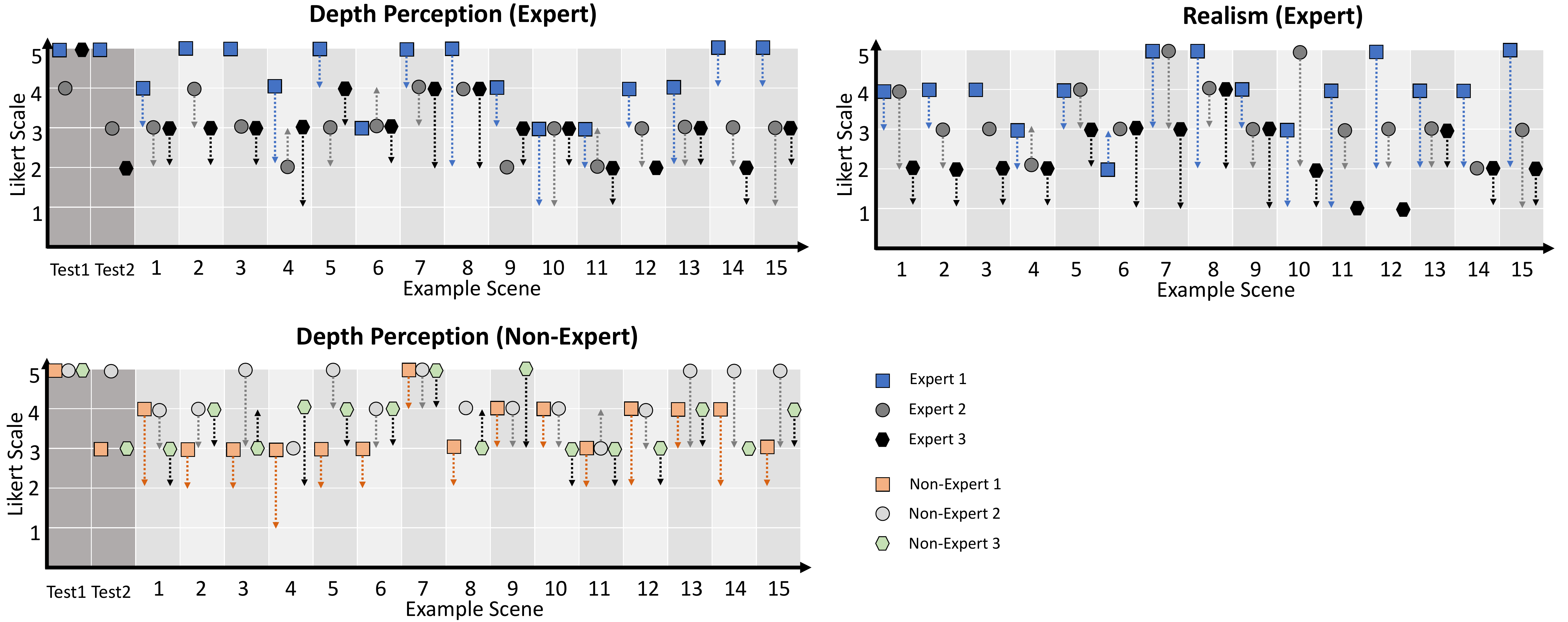}
   \end{center}
   \caption[] 
   { \label{fig:Diagr} Expert and non-expert ratings for depth perception and realism. Symbols indicate the rating per participant of the generated samples by cross-domain conditional GAN. Arrows show the difference to CycleGAN \cite{Zhu2017}.}
\end{figure}

Fig. \ref{fig:Diagr} illustrates the ratings by the non-experts for depth perception for each of the presented scenes. The median for each participant on our results in comparison to the baseline are $3$ to $2$, $4$ to $3$ and $3$ to $3$, which means that our method was clearly favored over the baseline and that the participants had a three-dimensional impression from the synthesized stereo images. Fig. \ref{fig:Diagr} also shows that depth perception even on real images of the silicone phantom is not assessed as completely perfect (\textit{Test1} and \textit{Test2}). These ratings help to relate the assessment of the generated stereo images to samples taken from the real world. 
In 39 instances, our method was preferred over the baseline by the participants (10 instances are better by $\Delta2$ and 29 are better by $\Delta1$). In three instances, both methods were assessed as equally good and in three other instances, our method was rated worse in stereo consistency. 

Considering the evaluation by the expert, a similar picture can be drawn. The respective diagram on assessment of depth perception is provided in Fig. \ref{fig:Diagr}. In 35 instances, our method was preferred over the baseline by the experts (1 instance better by $\Delta3$, 8 instances better by $\Delta2$ and 26 instances better by $\Delta1$). In seven cases, both methods were assessed as equal and in three other instances, our method was assessed as worse in comparison to the baseline.
When referring to the realism, our method is also superior, and was conceived as less artifact-prone and more related to an intraoperative scene. Fig. \ref{fig:Diagr} illustrates the ratings by the experts. In 37 cases, our method was preferred over the baseline by the experts (5 instances better by $\Delta3$, 10 instances better by $\Delta2$ and 22 instances better by $\Delta1$). 
Pathology assessment on the synthesized stereo frames yielded a good result. 37 of 45 correct decisions were made solely by watching the generated stereo pair. Furthermore, in 42 of 45 cases, the correct instrument that is shown in the scene, was named. Please note that in some instances, the instruments are only visible by a small margin on the actual clipped image. Moreover, motion artifacts complicated the assessment in 2 cases (example 13, 14). In 43 of 45 assessments, the right surgical phase has been identified by the participants. 


\section{Discussion}

To the best of our knowledge, the presented approach is the first to address stereo-endoscopic scene transformation for minimally-invasive surgical training. In this paper, we propose a novel cross domain conditioning GAN, which is superior in synthesizing consistent and more realistic stereo data in comparison to the unpaired CycleGAN approach \cite{Zhu2017}. Due to conditioning on a second image, which is drawn from the target domain (real or generated content), the network is also able to generate images with less artifacts and with more realistic color, heterogeneous textures, specularities and blood. 
The reliability of the generated samples was indirectly assessed by asking clinically relevant end points considering visible pathology, surgical instrument and surgical phase. We want to especially emphasize that almost all of the questions could be correctly answered with high confidence.
In general, we decided against the conduction of a Visual Turing Test, as some shape-related features in the scene (e.g.\ a personalized ring shape instead of a standard commercial ring) would have been easily identified by an expert surgeon. 


Future work includes usage of the presented approach together with depth sensing technologies which are currently not applicable during surgery due to sterilization restrictions. 
The acquired depth information can be leveraged as ground truth data for training disparity estimation models from transformed mono- or stereo-endoscopic images.


\bibliography{bibliography_anonym}

\begin{thebibliography}{10}
\providecommand{\url}[1]{#1}
\csname url@samestyle\endcsname
\providecommand{\newblock}{\relax}
\providecommand{\bibinfo}[2]{#2}
\providecommand{\BIBentrySTDinterwordspacing}{\spaceskip=0pt\relax}
\providecommand{\BIBentryALTinterwordstretchfactor}{4}
\providecommand{\BIBentryALTinterwordspacing}{\spaceskip=\fontdimen2\font plus
\BIBentryALTinterwordstretchfactor\fontdimen3\font minus
  \fontdimen4\font\relax}
\providecommand{\BIBforeignlanguage}[2]{{%
\expandafter\ifx\csname l@#1\endcsname\relax
\typeout{** WARNING: IEEEtran.bst: No hyphenation pattern has been}%
\typeout{** loaded for the language `#1'. Using the pattern for}%
\typeout{** the default language instead.}%
\else
\language=\csname l@#1\endcsname
\fi
#2}}
\providecommand{\BIBdecl}{\relax}
\BIBdecl

\bibitem{EngelhardtMICCAI}
S.~Engelhardt, R.~De~Simone, P.~M. Full, M.~Karck, and I.~Wolf, ``Improving
  surgical training phantoms by hyperrealism: Deep unpaired image-to-image
  translation from real surgeries,'' in \emph{Medical Image Computing and
  Computer Assisted Intervention -- MICCAI 2018}, 2018, pp. 747--755.

\bibitem{milgram_taxonomy_1994}
P.~Milgram and F.~Kishino, ``A taxonomy of mixed reality visual displays,''
  \emph{IEICE Trans Inf Syst}, vol.~77, no.~12, pp. 1321--1329, 1994.

\bibitem{Luengo2018SurRealES}
I.~Luengo, E.~Flouty, P.~Giataganas, P.~Wisanuvej, J.~Nehme, and D.~Stoyanov,
  ``Surreal: enhancing surgical simulation realism using style transfer,'' in
  \emph{British Machine Vision Conference 2018, {BMVC} 2018, Northumbria
  University, Newcastle, UK, September 3-6, 2018}, 2018, p. 116.

\bibitem{Zhu2017}
J.~Y. Zhu, T.~Park, P.~Isola, and A.~A. Efros, ``Unpaired image-to-image
  translation using cycle-consistent adversarial networks,'' in \emph{2017 IEEE
  International Conference on Computer Vision (ICCV)}, 2017, pp. 2242--2251.

\bibitem{Chen_2018_CVPR}
D.~Chen, L.~Yuan, J.~Liao, N.~Yu, and G.~Hua, ``Stereoscopic neural style
  transfer,'' in \emph{The IEEE Conference on Computer Vision and Pattern
  Recognition (CVPR)}, June 2018, pp. 6654--6663.

\bibitem{Mirza2014}
M.~Mirza and S.~Osindero, ``Conditional {Generative} {Adversarial} {Nets},''
  \emph{arXiv:1411.1784}, Nov. 2014.

\bibitem{Isola2016}
P.~Isola, J.-Y. Zhu, T.~Zhou, and A.~A. Efros, ``Image-to-{Image} {Translation}
  with {Conditional} {Adversarial} {Networks},'' \emph{arXiv:1611.07004}, Nov.
  2016.

\bibitem{Yi2017}
Z.~Yi, H.~Zhang, P.~Tan, and M.~Gong, ``{DualGAN: Unsupervised Dual Learning
  for Image-To-Image Translation},'' in \emph{{The IEEE International
  Conference on Computer Vision (ICCV)}}, Oct 2017, pp. 2868--2876.

\bibitem{Engelhardt_IJCARS}
S.~Engelhardt, S.~Sauerzapf, B.~Preim, M.~Karck, I.~Wolf, and R.~De~Simone,
  ``{Flexible and Comprehensive Patient-Specific Mitral Valve Silicone Models
  with Chordae Tendinae Made From 3D-Printable Molds},'' \emph{International
  Journal of Computer Assisted Radiology and Surgery (IPCAI Special Issue)},
  vol.~14, no.~7, 2019.

\bibitem{Engelhardt_ICVTS}
S.~Engelhardt, S.~Sauerzapf, A.~Brčić, M.~Karck, I.~Wolf, and R.~De~Simone,
  ``{Replicated mitral valve models from real patients offer training
  opportunities for minimally invasive mitral valve repair},'' \emph{Interact
  Cardiovasc Thorac Surg.}, 2019.

\end{thebibliography}
\bibliographystyle{IEEEtran}

\end{document}